\documentclass[%
 reprint,
 amsmath,amssymb,
 aps,
]{revtex4-2}

\usepackage{graphicx}
\usepackage{dcolumn}
\usepackage{bm}
\usepackage{multirow}
\usepackage{xcolor}
\usepackage{hyperref}

\begin{document}

\title{A Proof of Generalized Kramers-Pasternack Relation using Hyper-Radial Equation}

\author{Avoy Jana}
\affiliation{%
 Master in Science, Physics\\
 Indian Institute of Technology, Delhi
}%

\date{\today}

\begin{abstract}
\textbf{Abstract} We present a proof of the generalized Kramers-Pasternack relation using the hyper-radial equation approach. Following Kramers’ method, we manipulate the radial equation by multiplying it with an expression closely related to terms in the hyper-virial theorem. Through successive integrations by parts, we systematically derive the second Pasternack formula, extending its validity to arbitrary dimensions. This approach provides deeper insights into the algebraic structure of recurrence relations for diagonal radial matrix elements in quantum mechanics.\\

\end{abstract}

\maketitle

\section{Introduction}
There is extensive literature on $\langle r^k\rangle$ for quantum mechanical systems in non-relativistic (Refs.~\cite{Bockasten1974,Dehesa2021,CorderoSoto2009,Aptekarev2010,Gonzalez2017}) and relativistic (Refs.~\cite{Andrae1997,Suslov2009,Suslov2008}) regimes. However, they do not possess a simple form. Therefore, we use the so-called Pasternack two-term and three-term recurrence relations. There are numerous ways to prove the Pasternack recurrence relation, such as using the recurrence relation of the Laguerre polynomial by I. Waller (1926) (Ref.~\cite{Waller1926}), Dirac q-number by J. H. Van Vleck (1934) (Ref.~\cite{VanVleck1934}), generalized hypergeometric functions by Pasternack (1937) (Ref.~\cite{Pasternack1937}), manipulations on the radial equation by Kramers (1938) (Ref.~\cite{Kramers1938}) (an English translation appeared in late 1944, leading to both being equivalently credited as the Kramers-Pasternack relation), the Hyper-Virial theorem by J. H. Epstein and S. T. Epstein (1962) (Ref.~\cite{Epstein1962}), the group theoretical approach by L. Armstrong (1971) (Ref.~\cite{Armstrong1971}) and Schrodinger factorization method by Tomasz Szymanski, J. K. Freericks (2020) (Ref.~\cite{szymanski2020}), among others.
The two-term recurrence relation, also known as the first Pasternack relation or the Pasternack inversion relation, is given by
\begin{equation}
    \begin{aligned}
        &\langle n, l | \hat{r}^{-k-2} | n, l \rangle = \left( \frac{2Z}{n a_0} \right)^{2k+1} \\
        &\times \frac{(2l  - k)!}{(2l  + k + 1)!}
        \langle n, l | \hat{r}^{k-1} | n, l \rangle
    \end{aligned}
\end{equation}
for $0\leq k \leq 2\ell$, which is particularly useful for determining negative moments. However, for positive moments, we use the three-term recurrence relation, also known as the second Pasternack relation or the Kramers-Pasternack relation, given by
\begin{equation}
    \begin{aligned}
        &\frac{(k+1)Z^2}{n^2a_0^2} \langle r^k \rangle - \frac{Z(2k+1)}{a_0} \langle r^{k-1} \rangle\\
        &+ \frac{k}{4} \left[(2l+1)^2 - k^2\right]  \langle r^{k-2} \rangle = 0
    \end{aligned}
\end{equation}
for $k\geq1$. For both cases, the initial conditions are
\begin{equation}
    \langle r^{-1} \rangle = \frac{Z}{n^2a_0}, \quad \langle 1 \rangle =1.
\end{equation}
Many popular textbooks (Refs.~\cite{zettili2010,griffiths2004,fitts1999,banks2018}) include this three-term relation with just mentioned or left as an exercise. Here, we are particularly interested in the $d$-dimensional case.
A $d$-dimensional form of the two-term recurrence relation, following Pasternack's technique of regularized hypergeometric functions, can be rewritten as
\begin{equation}
    \begin{aligned}
        &\langle n, l | \hat{r}^{-k-2} | n, l \rangle = \left( \frac{2Z}{\left(n+\frac{d-3}{2}\right) a_0} \right)^{2k+1} \\
        &\times \frac{(2l + d - 3 - k)!}{(2l + d - 3 + k + 1)!}
        \langle n, l | \hat{r}^{k-1} | n, l \rangle
    \end{aligned}
\end{equation}
which is easily deduced from the three-dimensional form. However, the three-term recurrence relation does not follow in the usual manner.
These recurrence relations apply to the diagonal elements of the radial matrix. S. Pasternack, R. M. Sternheimer (1962) (Ref.~\cite{pasternack1962}), and Blanchard (1974) (Ref.~\cite{blanchard1974}), among others, worked on the non-diagonal elements of matrix elements. A useful generalized form of Blanchard's recurrence relation (Ref.~\cite{sun2015}) is given by
\begin{equation}
    \begin{aligned}
        &(E_{n_1 l_1} - E_{n_2 l_2})^2 \langle n_1 l_1 | r^k | n_2 l_2 \rangle = \xi \langle n_1 l_1 | r^{k-4} | n_2 l_2 \rangle\\
        &+ \left[ \frac{k - 1}{k - 2}(l_1 - l_2)(d - 2 + l_1 + l_2)(E_{n_1 l_1} - E_{n_2 l_2}) \right.\\
        &\left. - k(k - 1)(E_{n_1 l_1} + E_{n_2 l_2}) \right] \langle n_1 l_1 | r^{k-2} | n_2 l_2 \rangle\\
        &+ k \langle n_1 l_1 | V'(r) r^{k-1} | n_2 l_2 \rangle \\
        &+ 2k(k - 1) \langle n_1 l_1 | V(r) r^{k-2} | n_2 l_2 \rangle
    \end{aligned}
\end{equation}
where
\begin{equation}
    \begin{aligned}
        &\xi = \frac{-k}{4(k-2)}[(l_1 - l_2)^2 - (k - 2)^2]\\
        &\times (d - k + l_1 + l_2)(d - 4 + k + l_1 + l_2).
    \end{aligned}
\end{equation}
For diagonal elements, setting $ n_1 = n_2 = n $ and $ l_1 = l_2 = l $, Blanchard's recurrence relation reduces to Kramers' recurrence relation,
\begin{equation}
    \begin{aligned}
        &\frac{1}{4}(k - 2)(d - k + 2l)(d - 4 + k + 2l) \langle n l | r^{k-4} | n l \rangle\\
        &- 2(k - 1) E_n \langle n l | r^{k-2} | n l \rangle + \langle n l | V'(r) r^{k-1} | n l \rangle\\
        &+ 2(k - 1) \langle n l | V(r) r^{k-2} | n l \rangle = 0.
    \end{aligned}
\end{equation}
A convenient form of it can be written as
\begin{equation}
\begin{alignedat}{2}
    &\frac{k}{4}\left[(d-1)(d-3)+4\ell(\ell+d-2)-(k^2-1)\right]\langle r^{k-2}\rangle\\
    &- 2(k +1) E_n \langle r^{k}\rangle + \langle V'(r) r^{k+1} \rangle+ 2(k +1) \langle  V(r) r^{k} \rangle = 0\\
\end{alignedat}
\end{equation}
If we wish to recover this relation with $\mu$ and $\hbar$, then terms involving $E_n$, $V(r)$, and $V'(r)$ must be multiplied by $\mu\hbar^{-2}$.
\section{Derivation Using Hyper-Radial Equation}
The independent Schrodinger equation is given by
\begin{equation}
    \hat{H} \psi = \left(- \frac{\hbar^2}{2\mu} \nabla^2 + V(r)\right) \psi = E \psi
\end{equation}
where \(V(r)\) is the potential. For a spherically symmetric potential, the wave function separates as
\begin{equation}
    \psi(r, \Omega) = R_{n\ell}(r) Y^K_{\ell}(\Omega)
\end{equation}
where \(R_{n\ell}(r)\) is the radial part and \(Y^K_{\ell}(\Omega)\) are spherical harmonics. The Laplacian in spherical coordinates takes the form
\begin{equation}
    \nabla^2 = \nabla_r^2 - \frac{\hat{L}^2}{r^2 \hbar^2}
\end{equation}
with the radial Laplacian given by
\begin{equation}
    \nabla_r^2 = \frac{1}{r^{d-1}} \frac{d}{d r} \left( r^{d-1} \frac{d}{d r} \right)
\end{equation}
The angular momentum operator satisfies
\begin{equation}
    \hat{L}^2 Y^K_{\ell}(\Omega) = \ell (\ell + d - 2) \hbar^2 Y^K_{\ell}(\Omega)
\end{equation}
Rewriting the Hamiltonian in terms of the radial operator gives
\begin{equation}
    \hat{H} = \frac{\hat{p}_r^2}{2\mu} + \hat{V}_{\text{eff}}(r)
\end{equation}
where the radial momentum operator satisfies
\begin{equation}
    \hat{p}_r^2 = -\hbar^2 \left(\nabla_r^2 + \frac{(d-1)(d-3)}{4r^2}\right)
\end{equation}
The effective potential (Ref.~\cite{jana2025generalizedradialuncertaintyproduct}) is
\begin{equation}
    \hat{V}_{\text{eff}}(r) = V(r) + \frac{\hat{L}^2}{2\mu r^2} + \frac{\hbar^2}{2\mu} \frac{(d-1)(d-3)}{4r^2}
\end{equation}
Thus, the problem reduces to solving the radial Schrodinger equation with the effective potential, incorporating both the central potential and the centrifugal barrier. The radial Schrodinger equation can be written as
\begin{equation}
\left[-\frac{\hbar^2}{2\mu} \nabla_r^2 +V(r) + \frac{\ell(\ell + d-2)\hbar^2}{2\mu r^2}\right]R(r)= E R(r)
\end{equation}
Consider an arbitrary function which is independent of the involved
potential $V(r)$,
\begin{equation}
    f(r)=r^{\frac{d-1}{2}}R_{n\ell}(r)
\end{equation}
which gives us
\begin{equation}
\begin{alignedat}{2}
    &f''(r)=\frac{1}{4}r^{\frac{d-5}{2}}\left[(d-1)(d-3)R(r)+ \right.\\
    &\left. 4r^2\left[\frac{(d-1)R'(r)}{r}+R''(r)\right]\right]
\end{alignedat}
\end{equation}
and we have
\begin{equation}
    \nabla_r^2 R(r)=\frac{1}{r^{d-1}}\frac{\partial}{\partial{r}}\left(r^{d-1}\frac{\partial R(r)}{\partial{r}}\right)=\frac{(d-1)R'(r)}{r}+R''(r)
\end{equation}
By combining Eqs. (18–20), we can write
\begin{equation}
    \nabla_r^2 R(r)=\frac{1}{r^{\frac{d-1}{2}}}\left(f''(r)-\frac{(d-1)(d-3)}{4r^2}f(r)\right)
\end{equation}
Using Eqs. (18,21) in Eq. (17) and rearranging bit,
\begin{equation}
    \frac{f''}{f}=-\frac{2\mu}{\hbar^2}(E-V(r))+\left[\frac{(d-1)(d-3)}{4}+\ell(\ell+d-2)\right]\frac{1}{r^2}
\end{equation}
Let dive into some manipulations,
\begin{equation}
\begin{alignedat}{2}
    &\int_0^{\infty} fr^kf''dr=-\frac{2\mu}{\hbar^2}E\int_0^{\infty} fr^kfdr+\frac{2\mu}{\hbar^2}\int_0^{\infty} fr^k V(r)f dr\\
    &+\left[\frac{(d-1)(d-3)}{4}+\ell(\ell+d-2)\right]\int_0^{\infty} fr^{k-2}f dr
\end{alignedat}
\end{equation}
Using simple integration by parts,
\begin{equation}
    \int_0^{\infty} fr^kf''dr=-\int_0^{\infty} f'r^kf'dr-k\int_0^{\infty} fr^{k-1}f'dr
\end{equation}
and again using integration by parts,
\begin{equation}
    \int_0^{\infty} f'r^kf'dr=-\frac{2}{k+1} \int_0^{\infty} f'r^{k+1}f''dr
\end{equation}
for which
\begin{equation}
    \int_0^{\infty} fr^kf''dr=\frac{2}{k+1}\int_0^{\infty} f'r^{k+1}f''dr -k\int_0^{\infty} fr^{k-1}f'dr
\end{equation}
It's important to note that, the surface terms vanishes in every integration by parts as $f(r\to 0)=0$ and $f(r \to \infty)=f'(r\to \infty)=0$.
Now using Eq. (22),
\begin{equation}
\begin{alignedat}{2}
    &\int_0^{\infty} f'r^{k+1}f''dr=-\frac{2\mu}{\hbar^2}E\int_0^{\infty} f'r^{k+1}fdr\\
    &+\frac{2\mu}{\hbar^2}\int_0^{\infty} f'r^{k+1} V(r)f dr\\
    &+\left[\frac{(d-1)(d-3)}{4}+\ell(\ell+d-2)\right]\int_0^{\infty} f'r^{k-1}f dr\\
\end{alignedat}
\end{equation}
Let try to find a general formula for $\int_0^{\infty} f'r^{p}fdr$, using integration by parts,
\begin{equation}
    \int_0^{\infty} f'r^{p}fdr=p\int_0^{\infty} fr^{p-1}fdr-\int_0^{\infty} f'r^{p}fdr
\end{equation}
which gives
\begin{equation}
    \int_0^{\infty} f'r^{p}fdr=-\frac{p}{2} \int_0^{\infty} fr^{p-1}fdr
\end{equation}
Using it, Eq. (27) will becomes
\begin{equation}
\begin{alignedat}{2}
    &\int_0^{\infty} f'r^{k+1}f''dr=\frac{2\mu}{\hbar^2}E\frac{k+1}{2}\int_0^{\infty} fr^{k}fdr\\
    &+\frac{2\mu}{\hbar^2}\int_0^{\infty} fr^{k} V(r)f dr\\
    &-\frac{k-1}{2}\left[\frac{(d-1)(d-3)}{4}+\ell(\ell+d-2)\right]\int_0^{\infty} fr^{k-2}f dr\\
\end{alignedat}
\end{equation}
From Eq. (26),
\begin{equation}
\begin{alignedat}{2}
    &\int_0^{\infty} fr^{k+1}f''dr=\frac{2\mu}{\hbar^2}E\int_0^{\infty} fr^{k}fdr\\
    &+\frac{2\mu}{\hbar^2}\frac{2}{k+1}\int_0^{\infty} f'r^{k} V(r)f dr\\
    &-\left[\frac{k-1}{k+1}\left[\frac{2k}{k+1}\frac{(d-1)(d-3)}{4}+\ell(\ell+d-2)\right] \right.\\
    &\left. -\frac{k(k-1)}{2}\right]\int_0^{\infty} fr^{k-2}f dr\\
\end{alignedat}
\end{equation}
Now equating Eq. (23) and Eq. (31),
\begin{equation}
\begin{alignedat}{2}
    &\frac{2\mu}{\hbar^2}2E\int_0^{\infty} fr^{k}fdr\\
    &+\frac{2\mu}{\hbar^2}\left[\frac{2}{k+1}\int_0^{\infty} f'r^{k+1} V(r)f dr-\int_0^{\infty} fr^{k} V(r)f dr\right]\\
    &-\left[\frac{2k}{k+1}\left[\frac{(d-1)(d-3)}{4}+\ell(\ell+d-2)\right]-\frac{k(k-1)}{2}\right]\\
    &\int_0^{\infty} fr^{k-2}f dr=0\\
\end{alignedat}
\end{equation}
Now rewriting in good prescription,
\begin{equation}
\begin{alignedat}{2}
    &\frac{2\mu}{\hbar^2}2(k+1)E\langle r^{k}\rangle\\
    &+\frac{2\mu}{\hbar^2}\left[2\int_0^{\infty} f'r^{k+1} V(r)f dr-(k+1)\langle r^{k} V(r)\rangle\right]\\
    &-\left[2k\left[\frac{(d-1)(d-3)}{4}+\ell(\ell+d-2)\right]-\frac{k(k^2-1)}{2}\right]\langle r^{k-2}\rangle\\
\end{alignedat}
\end{equation}
Let take a look on the term $\int_0^{\infty} f'r^{k+1} V(r)f dr$a and using integration by parts as usual upon it,
\begin{equation}
\begin{alignedat}{2}
    &\int_0^{\infty} f'r^{k+1} V(r)f dr\\
    &=-\int_0^{\infty} (f r^{k+1} V'(r)f+(k+1) f r^{k} V(r)f)dr\\
    &-\int_0^{\infty} f'r^{k+1}V(r)fdr\\
\end{alignedat}
\end{equation}
which gives
\begin{equation}
    2\int_0^{\infty} f'r^{k+1} V(r)f dr=-\langle r^{k+1} V'(r) \rangle - (k+1) \langle r^{k} V(r) \rangle
\end{equation}
Finally from Eq. (33), we have can achieve generalized Kramers-Pasternack recurrence relation, is given as
\begin{equation}
\begin{alignedat}{2}
    &\frac{\mu}{\hbar^2}2(k+1)E\langle r^{k}\rangle+\frac{\mu}{\hbar^2}\left[-\langle r^{k+1} V'(r) \rangle-2(k+1)\langle r^{k} V(r)\rangle\right]\\
    &-\frac{k}{4}\left[(d-1)(d-3)+4\ell(\ell+d-2)-(k^2-1)\right]\langle r^{k-2}\rangle=0\\
\end{alignedat}
\end{equation}

\section{Conclusion}
For Hydrogen-like systems, we have the potential form,
\begin{equation}
    V(r)=-\frac{ZKe^2}{r}=-\frac{\hbar^2}{\mu}\frac{Z}{a_0r}
\end{equation}
where $a_0=\frac{\hbar^2}{\mu Ke^2}$ is Bohr radius. For $d$-dimensional Hydrogen-like systems, the energy eigenvalues (Ref.~\cite{jana2025generalizedradialuncertaintyproduct}) is given by
\begin{equation}
    E_n=-\frac{\hbar^2}{\mu} \frac{Z^2}{2\left(n+\frac{d-3}{2}\right)^2 a_0^2}
\end{equation}
Then the Kramers-Pasternack recurrence relation becomes 
\begin{equation}
\begin{alignedat}{2}
    &\frac{(k+1)Z^2}{\left(n+\frac{d-3}{2}\right)^2a_0^2}\langle r^{k}\rangle-\frac{Z}{a_0}(2k+1)\langle r^{k-1}\rangle\\
    &+\frac{k}{4}\left[(d-1)(d-3)+4\ell(\ell+d-2)-(k^2-1)\right]\langle r^{k-2}\rangle=0\\
\end{alignedat}
\end{equation}
Let start with the initial seeds
\begin{equation}
    \langle r^{-1} \rangle = \frac{Z}{\left(n+\frac{d-3}{2}\right)^2a_0} \text{ and } \langle 1 \rangle =1
\end{equation}
Setting $k=1$ in Eq. (39),
\begin{equation}
\begin{alignedat}{2}
    &\langle r \rangle =\frac{a_0}{Z} \left[\frac{3}{2}\left(n+\frac{d-3}{2}\right)^2 \right.\\
    &\left. -\frac{1}{8}[(d-1)(d-3)+4\ell(\ell+d-2)]\right]\\
\end{alignedat}
\end{equation}
which can be simplified to our known result (Ref.~\cite{jana2025generalizedradialuncertaintyproduct}),
\begin{equation}
    \langle \hat{r} \rangle= \frac{1}{4}\frac{a_{0}}{Z}[d^2+d(6n-2\ell-7)+2(3n^2-9n-\ell^2+2\ell+6)]
\end{equation}
Setting $k=2$ in Eq. (39),
\begin{equation}
\begin{alignedat}{2}
    &\langle \hat{r^2} \rangle =\frac{a_{0}^2}{Z^2} \left(n+\frac{d-3}{2}\right)^2\\
    &\left[\frac{5}{3}\frac{Z}{a_0}\langle r \rangle-\frac{1}{12}[(d-1)(d-3)+4\ell(\ell+d-2)]\right]\\
\end{alignedat}
\end{equation}
which can be simplified again to our known result (Ref.~\cite{jana2025generalizedradialuncertaintyproduct}),
\begin{equation*}
\begin{alignedat}{2}
   &\langle \hat{r}^2 \rangle=\frac{1}{8}.\frac{\left(n+\frac{d-3}{2}\right)^2a_0^2}{Z^2}[d^3+d^2(12n-6\ell-12)\\
   &+d(30n^2-6\ell^2-12n\ell-78n+30\ell+47)\\
   &+(20n^3-12n\ell^2-90n^2+18\ell^2+24n\ell+130n-36\ell-60)]\\
\end{alignedat}
\end{equation*}
While setting $d=3,2$, we can get the Kramers-Pasternack recurrence relation are given by
\begin{equation}
    \frac{(k+1)Z^2}{n^2a_0^2}\langle r^k \rangle - \frac{Z(2k+1)}{a_0}\langle r^{k-1} \rangle +\frac{k}{4} [(2\ell+1)^2-k^2]\langle r^{k-2}\rangle=0
\end{equation}
\begin{equation}
    \frac{(k+1)Z^2}{\left(n+\frac{1}{2}\right)^2a_0^2}\langle r^k \rangle - \frac{Z(2k+1)}{a_0}\langle r^{k-1} \rangle +k \left[|\ell|^2-\frac{k^2}{4}\right]\langle r^{k-2}\rangle=0
\end{equation}
Using two-term relation of Eq. (4) and three term relation (for positive moments) of Eq. (39), we can set up a three-term relation for negative moments also.

\begin{acknowledgments}
I sincerely express my gratitude to the Indian Institute of Technology (IIT) Delhi, for granting me valuable access to various journal platforms. I also extend my heartfelt appreciation to my classmates, Molla Suman Rahaman and Tamal Majumdar, for their constant encouragement and support throughout my work.
\end{acknowledgments}

\end{document}